# *In Silico* Derivation of HLA-Specific Alloreactivity Potential from Whole Exome Sequencing of Stem Cell Transplant Donors and Recipients: Understanding the Quantitative Immuno-biology of Allogeneic Transplantation

Max Jameson-Lee, [1] Vishal Koparde, [2] Phil Griffith, [1] Allison F. Scalora, [1] Juliana K. Sampson, [2] Haniya Khalid, [1] Nihar U. Sheth, [2] Michael Batalo, [1] Myrna G. Serrano, [2] Catherine H. Roberts,[1] Michael L. Hess,[3] Gregory A. Buck, [2] Michael C. Neale, [4] Masoud H. Manjili, [5] Amir A. Toor. [1]

Stem Cell Transplant Program, Massey Cancer Center; [1] The Center for the Study of Biological Complexity; [2] Department of Internal Medicine; [3] Department of Psychiatry and Statistical Genomics; [4] Department of Microbiology and Immunology; [5] Virginia Commonwealth University, Richmond, VA 23298

Address correspondence to, Amir A. Toor, MD; Bone Marrow Transplant Program, Massey Cancer Center, Virginia Commonwealth University, Richmond, VA 23298.

Email: atoor@vcu.edu, Phone: 804-628-2389

Running title: HLA-specific alloreactivity potential in transplantation

Abstract: 235, Words: 6422, Figures: 7, Tables: 3, Supplementary figures: 2, Supplementary tables: 2

**Abstract**

Donor T cell mediated graft vs. host effects may result from the aggregate alloreactivity to minor histocompatibility antigens (mHA) presented by the HLA in each donor-recipient pair (DRP) undergoing stem cell transplantation (SCT). Whole exome sequencing has demonstrated extensive nucleotide sequence variation in HLA-matched DRP. Non-synonymous single nucleotide polymorphisms (nsSNPs) in the GVH direction (polymorphisms present in recipient and absent in donor) were identified in 4 HLA-matched related and 5 unrelated DRP. The nucleotide sequence flanking each SNP was obtained utilizing the ANNOVAR software package. All possible nonameric-peptides encoded by the non-synonymous SNP were then interrogated *in-silico* for their likelihood to be presented by the HLA class I molecules in individual DRP, using the Immune-Epitope Database (IEDB) SMM algorithm. The IEDB-SMM algorithm predicted a median 18,396 peptides/DRP which bound HLA with an IC50 of <500nM, and 2254 peptides/DRP with an IC50 of <50nM. Unrelated donors generally had higher numbers of peptides presented by the HLA. A similarly large library of presented peptides was identified when the data was interrogated using the Net MHCPan algorithm. These peptides were uniformly distributed in the various organ systems. The bioinformatic algorithm presented here demonstrates that there may be a high level of minor histocompatibility antigen variation in HLA-matched individuals, constituting an HLA-specific alloreactivity potential. These data provide a possible explanation for how relatively minor adjustments in GVHD prophylaxis yield relatively similar outcomes in HLA matched and mismatched SCT recipients.

**Introduction**

Graft versus host disease (GVHD) is a major impediment in achieving optimal outcomes in patients undergoing allogeneic stem cell transplantation (SCT) from human leukocyte antigen (HLA) identical related and unrelated donors.[1,2,3] Further, it remains unclear why with only relatively minor variation in GVHD prophylaxis, some patients with HLA-matched donors develop severe GVHD, whilst others with HLA-mismatched donors may not experience any.[4,5,6] In HLA matched donor-recipient pairs (DRP), a major contributor to GVHD occurrence are the peptides encoded by loci outside the major histocompatibility (MHC) locus on chromosome 6. These peptides, functionally defined as minor histocompatibility antigens (mHA), are presented by HLA molecules and are responsible for initiating both clinically beneficial graft versus tumor, and GVH responses.[7,8,9,10] As of 2012, around 49 mHA recognized by CD4+ or CD8+ T lymphocytes have been described.[11] Known mHA however, have a relatively low population distribution, particularly when the global populations are considered.[12] This fact, combined with the lack of genetic uniformity between individuals has limited discovery of mHA suitable for population-based screening before SCT.[13] Further complicating this problem is the HLA specificity of various mHA, and the heterogeneity observed in the HLA distribution in various populations across the world. Therefore in order to understand the biology and role of mHA in generating GVHD it is critical to quantify the extent of genetic variation between individuals.

Exploring genetic variation outside the MHC locus is also important to understand why, with relatively simple adjustments to the treatment protocols patients successfully engraft when transplanted with HLA mismatched donors. This is true for both unrelated donor umbilical cord blood transplant, and related haplo-identical SCT.[6] Moreover completely HLA-mismatched solid organ transplants result in successful engraftment, albeit with low-level life-long immunosuppression. Furthermore, organs such as kidney and heart tissues are prone to rejection when transplanted; yet, these organs are seldom targeted in GVHD, even in its chronic form, which affects nearly all organ systems. This makes it imperative to understand the role of mHA in generating alloreactivity, and the extent to which the magnitude of genetic variation outside the MHC locus contributes to allograft complications such as GVHD or graft rejection.

To examine these quantitative relationships, whole exome sequencing of SCT donor and recipients genomes was performed to measure the antigenic variability existing between them.[14] A large number of single nucleotide polymorphisms (SNP) were identified between donors and recipients.  These differences were classified as, either possessing, a GVH vector, polymorphisms present at loci in the recipient and absent in the donor, or, a HVG vector, present in the donor and absent in the recipient. The large number of SNPs in the exome, termed *alloreactivity potential*, suggests that in all individuals undergoing SCT there is a very high probability of there being peptides which may function as mHA. However, given the observed frequency of GVHD, seemingly, not all of these SNPs would lead to immunogenic peptides being generated, to yield clinically relevant mHA responses. This may be because, for HLA class I molecules on an antigen-presenting cell to present a peptide to an effector T lymphocyte, first, the endogenous protein must be cleaved by the proteasome, then the resulting peptides must bind HLA class I molecules to be presented. This would initiate either an immune response or result in tolerance, depending on the cellular and cytokine milieu at the time of antigen presentation.[15]

It is possible to determine the genetic variation between SCT recipients and donors, and to then bioinformatically determine the amino acid sequence of peptides resulting from SNPs encountered in their exomes. Further, bioinformatic techniques have been developed to determine which peptide antigens may be presented by specific HLA molecules. The Immune Epitope Database (IEDB; http://www.iedb.org) has characterized hundreds of thousands of peptides that can bind several hundred MHC complexes. From this large dataset, researchers have developed tools to predict peptide-HLA binding probabilities.[16] Initially, matrix-based methods such as stabilized matrix method (SMM)[17] were developed to determine binding affinities. More recently, neural network based algorithms such as NetMHC can use binding information from neighboring residues to predict dissociation constants between HLA molecules and putative mHA.[18] Finally, "pan-specific" algorithms have developed that are able to predict peptide-binding HLA alleles with limited experimental binding data.[19]

In this paper, the putative mHA in HLA-matched DRP and the *in silico* determined HLA class I binding affinity of these peptides is explored utilizing a bioinformatic approach based on exome sequencing of donors and recipients of SCT. The algorithm developed, lays a framework for future analysis of large SCT patient cohorts, and defines a personalized *HLA-specific alloreactivity potential*. The alloreactivity potential concept is analogous to the idea of potential energy in physics, i.e. the stored energy in a system. Thus HLA-specific alloreactivity potential would give an estimate of the likelihood that GVHD or graft rejection may develop in an HLA matched DRP in the absence of immunosuppression. Our work demonstrates that, the number of potentially immunogenic peptides varies considerably across HLA-matched related and unrelated donors, constituting a large alloreactivity potential.

**Methods**

*Whole exome sequencing*

Patients with recurrent hematological malignancies enrolled in a Virginia Commonwealth University Institutional Review Board approved protocol (Clinicaltrial.gov identifier: NCT00709592) were included in this study. To identify all the potentially immunogenic differences that exist in a SCT DRP, whole exome sequencing was performed on previously cryopreserved DNA from the donors and recipients enrolled in this study as previously described.[14] Of the nine DRP examined, 4 were from HLA-A, B, C and DRB1 matched related (MRD) and 5 from unrelated donors (URD). Histocompatibility testing was performed using high resolution typing for both HLA-class I (Table 1) and HLA-class II loci (*not shown*). The whole exome sequence of individual donors and recipients was compared both within pairs, and to a reference genome to identify all the SNPs, which were subsequently characterized as either synonymous or nonsynonymous. Next, all the nonsynonymous SNP (nsSNP) present in the recipient, but absent in the donor were identified, and designated as possessing a graft versus host (GVH) vector ($nsSNP_{GVH}$).

*Deriving HLA-specific alloreactivity potential*

To derive the amino acid sequence of the oligopeptides, i.e. potential mHA, resulting from these nsSNPs and their binding affinity to the relevant HLA in each DRP, a bioinformatics pipeline was developed. This pipeline has the following components: (1) determine $nsSNP_{GVH}$ between the exomes of transplant donors and recipients; (2) generate putative immunogenic peptides *in silico* from these genomic differences; and (3) analyze the binding affinity of these polymorphic peptides to the HLA in that individual (Figure 1). This third step estimates the likelihood of these peptides to be presented by the six patient-specific HLA class I molecules to determine *candidate* mHA. A complete description of this bioinformatic pipeline follows.

*Creation of peptide libraries:*

All the nsSNP$_{GVH}$ for each DRP were exported as variant call files (VCF) to the ANNOVAR software package. [20] Next, using the DB SNP130 database and hg18 genome coordinates of the nsSNP$_{GVH}$, amino acid sequences of the putative peptides were generated using the 'seq_padding' option of the 'annotate_variation' function in ANNOVAR. Endogenous peptides are presented by HLA class I molecules, and the average length of peptides binding HLA class I is 9 amino acids. Therefore for each polymorphism ANNOVAR returned 8 amino acids on either side of the nsSNP$_{GVH}$-encoded amino acid, resulting in a 17-mer peptide. This effectively generated nine nonamers from each nsSNP$_{GVH}$-encoded polymorphism; thus, the resulting peptides would have the polymorphic amino acid at positions 1 through 9, from the C to the N-terminal position (Figure 1).

*In silico variant peptide-HLA binding affinity determination*

The 17-mer peptides generated by ANNOVAR resulting from the nsSNP$_{GVH}$ were analyzed by the IEDB-MHC I-peptide binding prediction tools version 2.9.1, downloaded from (http://tools.immuneepitope.org/analyze/html_mhcibinding20090901B/download_mhc_I_binding.html). Nine oligopeptides were created for each 17-mer peptide using a 9-mer sliding window. The binding affinity of each of these 9-mers to the patient-specific HLA-A, HLA-B, and HLA-C (Table 1) were determined by running each 9-mer independently through the IEDB-MHC I prediction software. The output of this iterative process included variables such as, the gene name and coordinates, the polymorphic peptide sequence, and the calculated IC50 value via the SMM algorithm (a partial example of output in Supplementary table 1). IC50 values in nano-Molar (nM) represent the concentration of the test peptide, which will displace 50% of a standard peptide from the HLA molecule in question. The lower the IC50 for a peptide, the stronger the binding affinity of that peptide for the HLA in question. The cut-off in our analysis to classify a putative peptide as being *presented* by HLA, is an IC50 of < 500 nM. Those peptides that bound to HLA with an IC50 of < 50nM were designated *strongly presented*.

To validate the findings from the SMM algorithm, the ANNOVAR generated 17-mer peptide libraries were next interrogated using the NetMHCpan software

(http://www.cbs.dtu.dk/services/NetMHCpan/). To accomplish this, 2 software programs were developed to analyze the peptide data and query NetMHCpan remotely. The first program sequentially sent packets of 30 protein sequences to NetMHCpan. The protein sequences were sent in order by patient and HLA, and a sliding 9-mer window was selected to interrogate HLA binding, similar to SMM IEDB algorithm. NetMHC then returned *html* results, which were then stored on the local server. The second program examined the returned *html* results and organized it in a comma-separated-value (.csv) file, which could then be opened in Microsoft Excel for further analysis.

Results from the SMM IEDB algorithm and NetMHCpan were compared in each DRP by HLA loci and polymorphic peptides. Specifically, HLA locus and polymorphic peptide were combined to make a single variable within each patient dataset, allowing for the removal of duplicate peptides and identification of unique polymorphic peptides found by both or one methods. Presented and strongly presented polymorphic peptides were compared between the two methods, and then combined to get a comprehensive list of unique polymorphic peptide-HLA complexes for each patient.

*Deriving HLA-specific alloreactivity potential*

Given the large number of peptides strongly binding HLA identified in each DRP, area under the curve for the IC50 of the strongly binding peptides was determined to summarize the data. The peptide-HLA IC50s were plotted in an ascending order (descending order of affinity). First the non-linear distribution function of the peptides up to an IC50 of 100 mM was computed (a polynomial function of the second order). To obtain the area under the curve depicting the peptide-HLA complexes and their corresponding dissociation constants, the definite integral of the curve was determined. The definite integral by definition is the area of the *xy* plane bounded by the curve (1),

$$\int_a^b f(x)\,dx \tag{1}$$

Where *f(x)* denotes the function of the curve and a and b are the bounds on the x-axis, i.e. the lowest value of the IC50 recorded and the cutoff chosen.

*Tissue expression of polymorphic peptides*

Relative gene (and protein) expression level is a critical factor contributing to HLA Class I presentation of a peptide derived from the gene. [21] To investigate the tissue distribution of the *presented* peptides, software from the European Bioinformatics Institute, Illumina Body Map, (http://www.ebi.ac.uk/arrayexpress/experiments/E-MTAB-513/) was used to correlate *presented* peptides from the peptide library with relative gene expression in different tissues.

**Results**

*Creation of polymorphic peptides*

Whole exomes of 9 SCT DRP were sequenced, identifying an average of 6,445 nsSNP between donors and recipients. To determine the nsSNP that would constitute possible mHA, peptide sequences were generated that incorporated the polymorphic amino acid at each position 1 thru 9, in a nonameric peptide using the ANNOVAR software. Theoretically this could yield nine different peptides, however a nsSNP near either the 3' or 5' end of a sequence of a gene (N or C terminus of a protein) would lead to fewer peptides. The ANNOVAR output yielded on average 486,463 potential peptides encoded by nsSNPs and presented by the six HLA molecules (Range: 1,043,514-366,426 peptides/DRP). This output was generally greater than the calculated possibilities since it also included peptides resulting from splice variants of the various proteins bearing SNP encoded amino acids. In all, these peptides constituted the total pool of variant peptides, which may be immunogenic in a DRP (Figure 2).

*HLA-specific alloreactivity potential*

The 9-mer peptides bearing the polymorphic peptide in the GVH direction were then analyzed for their binding affinities to the individual HLA class I in each patient to determine the peptides potentially presented to the donor T cells. The IEDB SMM HLA class I binding prediction algorithm was utilized to calculate the binding affinity of the peptide output from ANNOVAR, and to rank putative mHA for their ability to be presented by individual HLA. After filtering for splice variants and duplicate peptide representation in the data set, this yielded a median of 18,396 (range, 1,926-72,294) peptides that bound with an IC50 of < 500 nM to HLA-A, -B and -C in the nine DRP, and were designated as *presented*. Further, a median number of 2,254 (177-21,548) peptides were predicted to bind MHC class I with an IC50 of < 50 nM and were designated as *strongly presented* (Figure 2). When separated by the donor type (MRD, n=4, vs. URD, n=5), the HLA-matched unrelated DRPs had a significantly higher number of both *presented* and *strongly presented* peptides as determined by IEDB SMM (P=0.016 Mann-Whitney U-test) (Figure 3). The difference in the number of *presented* peptides between unrelated and related donors corroborated the large alloreactivity potential identified earlier in these donor types by whole exome sequencing.[14] Despite this there was no correlation identified between the number of SNPs and the number of *presented* peptides.

To summarize the mass of information on the peptides and their binding affinities, the peptides were ranked according to their IC50 and the distribution of their binding affinities was determined (Figure 4). This operation was performed without filtering duplicate peptide-HLA complexes resulting from splice variants. Area under the curve for each DRP was then computed for peptides with an IC50 up to 100nM. Once again, marked differences were observed in the AUC between matched related and unrelated donors (Table 2). This summarized measure hypothetically represents an HLA-specific alloreactivity potential for each unique DRP, and may be considered as an example of the cumulative mHA differences observed between the HLA matched donors and recipients.

In a further analysis, when the reciprocal of the IC50 for each peptide (a measure reflecting binding affinity) was plotted for each peptide, a Power distribution was observed, analogous to T cell clonal frequency distribution previously reported (Supplementary Figure 1).[22]

*Verifying HLA binding affinity of the variant peptide library in unique DRP*

To confirm the IEDB SMM algorithm findings, a second peptide-HLA binding affinity prediction tool, NetMHCpan, was used to interrogate the variant peptide libraries from the unique DRP and its output compared with the IEDB SMM. The NetMHCpan yielded a median of 3,962 peptides categorized as *presented* and 989 peptides as *strongly presented* (MRD vs. URD, P=0.063 & 0.11 respectively, Mann-Whitney U test) (Table 3). The IEDB-SMM and NetMHCpan data sets were then combined and unique peptide-HLA complexes predicted to be presented by both algorithms determined (*shared* peptides). The median number of *shared* unique peptides presented/DRP was 2,065 (range, 417-4,881) (Table 3). A representative data table depicting peptide sequences and respective IC50 values for binding to a single HLA locus, in a patient, predicted by both algorithms is given in Supplementary table 1. Plotting the IC50 of unique *presented* peptide-HLA complexes derived utilizing both algorithms, demonstrated not only a very large number of complexes, but also that a large proportion of these complexes were categorized as *strongly presented* (Figure 5). Furthermore, a weak, but significant correlation was identified between the IC50 predictions for both the algorithms in the shared peptide-HLA complex data sets (N=9, median Pearson's correlation coefficient R=0.62, P<0.01). Additionally, when the distribution of peptides presented on the three class I HLA loci was examined, no discernable preference for particular HLA loci was observed in terms of likelihood of peptide presentation (Supplementary Figure 2A &2B), except for a possible HLA-C dominance in URD recipients in the SMM algorithm.

*Tissue distribution of peptides*

For a peptide to be relevant in terms of its contribution to GVHD risk, in addition to its potential for presentation on the relevant HLA in a specific DRP, the relevant protein needs to be expressed in the tissues. When the tissue distribution of the genes with putative mHA (presented peptides, IC50 <500 nM) was examined, a relatively uniform distribution was observed (Figure 6). Further, although several antigens are expressed in organs such as, colon, liver and lungs, frequent target organs for GVHD; a large number of genes bearing potentially antigenic peptides are also expressed in other organ systems such as the kidney and adipose tissue seldom targeted by GVHD (Supplementary table 2).

**Discussion**

Allogeneic SCT represents a unique model system to study donor T cell responses to neo-antigens encountered in the recipient. Unlike experimental animal models, however, clinical transplantation is characterized by a vast repertoire of variant antigens, which in theory would result in a complex expansion of the T cell repertoire. [23, 24] The findings reported here provide a direct estimate of the antigenic variation, which may be encountered by the donor cytotoxic T cell populations following SCT. Starting from nsSNPs in the exome of donors and recipients, the reported analysis determined the resulting variant nonameric peptides and gave an *in silico* estimate of the binding affinity (IC50) of these peptides to the relevant HLA in the transplant recipients. The existence of this very large library of immunogenic peptides in HLA-matched DRP, immediately raises the question as to why only some and not all the patients develop GVHD.

If all the peptides in this large library of potential mHA, were presented to non-tolerant T cells then GVHD would potentially develop in all SCT patients, particularly with unrelated donors, where this magnitude is considerably larger than MRD. Supporting this notion is the observation that development of extensive chronic GVHD in patients is relatively common when conventional immunosuppressive regimens are used. Further, our findings offer a possible explanation for why most patients develop GVHD despite having HLA identical donors, and do so more frequently when the donors are unrelated. [25,26] Alternatively, the large magnitude of mHA between HLA matched donors also gives an insight into why patients undergoing HLA mismatched transplants such as haplo-identical or mismatched unrelated donor transplants have clinical outcomes which are not dramatically different from those of HLA matched related donors, that is, if appropriate GVHD prophylaxis is used in the first few weeks of the transplant. [27, 28] This paradox may be understood, if one considers the mHA as the *targets* for GVHD and HLA as the *mediators* of this phenomenon. Thus, if the number of *targets* is relatively similar in HLA-matched and haplo-identical-related donor, and in the HLA-matched and -mismatched unrelated donor transplant recipients; the difference introduced by HLA mismatching is overcome by adjustments in the GVHD prophylaxis regimens. Thus, even though thousands of immunogenic peptides are present, the conditions at the time of transplantation determine eventual outcome following transplant, that is, whether tolerance will develop or GVHD ensue following the initial interaction between recipient mHA-HLA complexes and donor T cells. As an example, when the proteasome inhibitor bortezomib is added to the conditioning regimen, by inhibiting peptide generation, and consequently diminishing antigen presentation to donor T cells in the very first weeks of the transplant, it reduces the risk of GVHD in unrelated donor SCT. [6]

If the model outlined above is correct, then the enormous magnitude of immunogenic peptides constituting the HLA specific alloreactivity potential will constitute an antigenic 'pressure' upon the non-tolerant donor T cells when first encountered, influencing the evolving T cell repertoire following SCT. This antigenic pressure may be mitigated by agents, which influence either antigen presentation (e.g. bortezomib) or the T cell response (e.g. anti-thymocyte globulin, calcineurin inhibitors, mycophenolate mofetil, post-transplant cyclophosphamide). An observation from this data set that supports this hypothesis, is that the frequency distribution of the binding affinities of the peptides to the HLA molecules follows the Power law (Supplemental Figure 1). This frequency distribution is similar to the T cell clonal frequency distribution observed when T cell clonality is measured using high-throughput T cell receptor $\beta$ sequencing. [22] This suggests that the T cell repertoire and clonal frequency emerging after SCT may be proportional to the antigenic peptide-HLA binding affinities. Thus, peptides strongly bound to the HLA will elicit a strong T cell clonal response, if they engage a T cell receptor and appropriate co-stimulation is provided. And since the peptide antigen binding affinities form a continuum, rather than discrete clusters of high and low affinity, the T cell repertoire frequency similarly forms a continuum, described by the Power law. Another conclusion to be considered from the non-discrete distribution of peptide-HLA binding affinity is that other non-recipient derived antigens, such as pathogen-associated peptides may also lie on this continuum. This may result in *cross-reactivity* between autologous antigens and pathogen-associated peptides. [29] A manifestation of this in the transplant setting is the triggering of GVHD or graft rejection events by viral infections, such as cytomegalovirus or human herpes virus 6 virus infections. [30,31,32]

Can these findings be used to develop a clinically relevant model for allogeneic SCT? One possible explanation of the variant outcomes following SCT is that post-transplant emergent T cell clones either develop tolerance to the many antigens encountered or fail to do so depending on the milieu encountered in the host. Early interventions, such as administration of anti-thymocyte globulin,[33] bortezomib or post transplant cyclophosphamide have a large impact on late post transplant outcomes. Similar tolerance induction is observed following cellular interventions such as regulatory T cell infusion and conditioning which up regulates NK-T cells at the time of SCT. [34,35] This suggest that if a large antigenic pressure from the HLA-specific alloreactivity potential exists in all patients, then tissue injury and cytokine milieu at the time of SCT are critical in determining the development of GVHD. Thus when there is tissue injury following SCT, multiple antigens are presented, then in the absence of adequate immunosuppression, the T cell repertoire that develops results in the development of GVHD. On the other hand, if tissue injury is minimized and there is adequate immunosuppression, when the initial T cell-antigen presenting cell interactions take place, peripheral (or central) tolerance would emerge. Following that depending on the presence or absence of thymic tissue, T cell clones developing from infused stem cells may perpetuate this process based on the prevailing T cell population and the state of tissue injury (Figure 7). In such a model inflammation provoked by the acute GVHD initiated by infused donor-derived T cells reacting to recipient antigens is perpetuated in the form of 'auto-reactivity' by the T cells, developing from infused stem cells in the absence of normal thymic processing. This concept may not be novel in itself, however our model provides the first biologically plausible explanation reconciling mHA differences observed in HLA matched DRP.

The immunogenic peptides appear to be uniformly distributed in the major organ systems of the body. This raises the following question; why do solid organ transplant recipients develop rejection, but GVHD does not commonly affect most such organs, such as the kidney and heart? The data presented in this paper suggests a possible answer to this question if the above quantitative model of immunobiology of transplantation is considered. Thus in the days following SCT, when the infused donor T cells encounter widespread variant immunogenic recipient antigens in *inflamed* tissues with a large tissue interface for T cell-antigen presenting cell interaction, i.e, skin, GI mucosa, liver and lungs, there is a corresponding polyclonal T cell allo-immune response which may result in GVHD affecting the targeted organs. In contrast, the relatively smaller tissue interface in the absence of direct injury, in organs such as the heart and kidney, do not trigger an immunogenic response in the face of an ongoing, *competing* oligoclonal T cell response elicited by the larger organ systems with injury. When solid organ transplantation is performed, tissue injury in the transplanted organ resulting from the transplant procedure serves as the injury stimulus triggering graft rejection. Based on these data a mathematical model has been proposed to explain the notion of alloreactivity potential and its relationship with GVHD onset and propagation over time as in a 'chaotic dynamical system'. [36]

A potential therapeutic application of this analysis would be the ability to 'titrate' the intensity of immunosuppressive therapy in the peri-transplant period based on the magnitude of the HLA-specific alloreactivity potential. This study supports the need for intensive immunosuppression in patients undergoing unrelated donor allogeneic SCT, making this algorithm a useful analysis for treatment planning. [37] For example, if a patient has a high number of predicted mHA and these are over-represented in lung tissue, therapies can be specifically tailored for that patient and symptoms of lung GVHD treated more promptly. However, large-scale protein expression studies by Ponten et al concluded that most proteins are expressed in most tissues – "different tissues acquire their unique characteristics by controlling not which proteins are expressed but how much of each is produced". [38] This raises the question of which parameter plays a larger role in peptide presentation by MHC class I HLA: the absolute molar amount of protein expressed in a tissue, or the binding affinity for a particular peptide; our data suggests that it may be a combination of the two (Figure 7).

Aside from providing insights into the immunopathology of SCT, this work also highlights the potential of next generation sequencing for therapeutic application. Examination of donor-recipient and tumor exomes may allow identification of tumor specific peptides generated by mutations present in the tumor exome, but absent in the donor and recipient genome. Translation of these into peptides with high binding affinity to the HLA *in silico* may provide the opportunity to inexpensively generate panels of multiple tumor specific peptide-vaccines, which may be used to vaccinate the recipient following SCT. NetMHC has been used in conjunction with exome sequencing to quickly identify mHA-encoded, tumor-specific antigens in individual patients that were recognized by bulk tumor-infiltrating lymphocytes. [39] Our work provides the proof of principle that sequencing along with bioinformatics techniques may pave the way for this advance in immunotherapy of hematological malignancies.

As with any *in silico* work, this work can only be considered preliminary and the peptide-HLA class I combinations predicted in our work, will need experimental verification. Acknowledging this limitation, it should be noted that the accuracy of these algorithms has been reviewed and they have been found to be useful predictors of HLA presentation. As an example, in a vaccinia virus challenge mouse model, the NetMHC algorithm was able to predict epitopes responsible for 95% of the cytotoxic T cell (CTL) response with an IC50 threshold of < 500nM. [40] Similarly, Armistead et al. found that with an IC50 threshold of < 500nM, all peptides predicted by SMM-IEDB algorithm bound HLA-A 0201 in their assays. [41] To put our data in context, Nivjeen et al. created a database from all known nsSNPs that had been deposited in NCBI's dbSNP database which is labeled as all possible mHA in humans in Figure 2. [42, 43] In light of these findings it is not at all surprising that we find a large library of immunogenic mHA in each DRP, and there may exist a similar alloreactivity potential mediated by HLA class II.

In conclusion, the findings reported here demonstrate that whole exome sequencing, followed by *in silico* peptide generation and HLA binding affinity determination reveal a large and previously unmeasured *HLA-specific* alloreactivity potential. This potential is predictably larger in patients undergoing unrelated donor SCT and mirrors previously described T cell clonal frequency distribution. This large alloreactivity potential in HLA matched DRP, validates the notion that cellular and cytokine milieu at the time of transplantation is critical in determining transplant outcomes. In doing so it gives a plausible quantitative biological explanation of the relative ease with which transplants from alternative donors have become established as safe therapies. We posit that these methodologies may be used to develop mathematical models to better understand the immunopathology of SCT from both HLA matched and mismatched donors and may in the future allow more precise titration of the immunosuppression intensity in transplant recipients.

Acknowledgements: *The authors gratefully acknowledge the Massey Cancer Center Pilot Project Grant, JUP 11-0.9, and Virginia's Commonwealth Health Research Board Award #236-11-13 for funding.* The authors also gratefully acknowledge Dr. Jamie Teer (Moffit Cancer Center, Tampa, FL) for his helpful suggestions, critical in determining peptide sequence from exome sequence variation.

**Table 1. HLA typing of the donor-recipient pairs.** Patients 2, 4, 16 and 23 underwent MRD and the others URD SCT. Patient 2 had a single locus HLA-B antigen mismatch; patients 3, 7 and 10 had a male donor/female recipient combination and others were gender-matched.

| D-R Pair | HLA-A | HLA-A | HLA-B | HLA-B | HLA-C | HLA-C |
|---|---|---|---|---|---|---|
| 2 | 02:01 | 24:02 | 15:16 | 27:05 | 02:02 | 17:01 |
| 3 | 03:01 | 11:01 | 07:02 | 55:01 | 03:03 | 07:02 |
| 4 | 23:01 | 30:02 | 15:03 | 44:03 | 02:10 | 07:18 |
| 5 | 01:01 | 03:01 | 570101 | 07:02 | 07:02 | 07:01 |
| 7 | 01:01 | 02:01 | 44:02 | 55:01 | 03:03 | 05:01 |
| 8 | 01:01 | 24:02 | 07:02 | 55:01 | 03:03 | 07:02 |
| 10 | 01:01 | 03:01 | 080101 | 40:01 | 03:04 | 07:01 |
| 16 | 01:01 | 26:01 | 13:02 | 27:05 | 02:02 | 06:02 |
| 23 | 03:01 | 24:02 | 07:02 | 57:01 | 06:02 | 07:02 |

**Table 2: HLA Specific Alloreactivity potential.** Area under the curve for the peptide IC50, calculated for all the peptides with an SMM-IC50 of <100 nM (algorithm output not filtered for duplicate peptides/splice variants). This value summarizes the number of peptides with a high binding affinity and their binding affinities (See Figure 4). Unrelated DRP are shaded gray.

| Patient | AUC (NM.Peptide) |
|---|---|
| 2 | $0.0361*10^6$ |
| 4 | $0.1191*10^6$ |
| 16 | $0.0417*10^6$ |
| 23 | $0.1906*10^6$ |
| 3 | $2.5802*10^6$ |
| 5 | $0.4751*10^6$ |
| 7 | $2.2249*10^6$ |
| 8 | $1.9886*10^6$ |
| 10 | $0.3754*10^6$ |

**Table 3: Number presented and strongly presented peptides predicted by the IEDB SMM & NetMHCPan algorithms**. Last column represents number of unique peptides predicted to bind the relevant HLA by both algorithms. Unrelated DRP are shaded gray. Presented and strongly presented peptide-HLA complexes have IC50 of <500 and <50 nM respectively.

| Patient | nsSNP$_{GVH}$ | SMM *Presented* | SMM *strongly presented* | NetMHC *presented* | NetMHC *strongly presented* | Shared peptides *presented* |
|---|---|---|---|---|---|---|
| 2 | 4,446 | 1,926 | 250 | 3,883 | 1,376 | 1,332 |
| 4 | 4,448 | 5,412 | 825 | 3,962 | 885 | 2,441 |
| 16 | 3,290 | 2,111 | 177 | 1,071 | 427 | 417 |
| 23 | 3,657 | 13,456 | 705 | 787 | 118 | 534 |
| 3 | 7,227 | 72,294 | 21,339 | 7,242 | 2,509 | 4,881 |
| 5 | 6,572 | 30,730 | 2,254 | 2,759 | 538 | 1,865 |
| 7 | 6,725 | 58,209 | 21,548 | 5,231 | 2,178 | 2,931 |
| 8 | 6,573 | 65,298 | 19,275 | 4,831 | 2,000 | 2,445 |
| 10 | 9,203 | 18,396 | 2,283 | 5,002 | 989 | 2,065 |

**Figure 1: Bioinformatics Workflow for calculating HLA specific alloreactivity potential in individual DRP**. Starting with donor and recipient whole exome sequence data, $nsSNP_{GVH}$ were identified, and peptide fragments generated using the ANNOVAR software package. These peptides, together with HLA data (Table 1) were then analyzed with IEDB SMM and NetMHCpan algorithms separately. Individual DRP binding data was then analyzed and candidate mHAs catalogued.

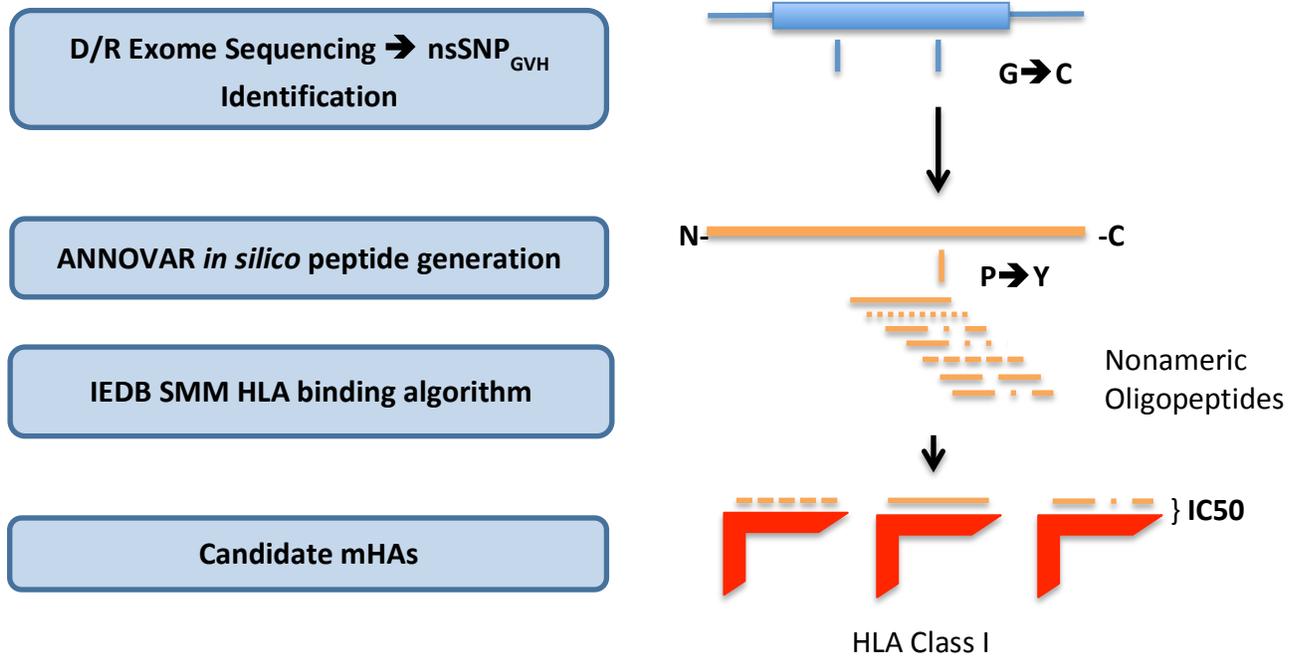

**Figure 2: The burden of minor histo-incompatibility in human SCT. A) All possible mHA in humans**: Data generated from NCBI dbSNP database.[42] **B) Alloreactivity potential:** The current patient cohort had an average of 6,445 nsSNPs/DRP, which when converted into peptide fragments averaged 486,463 possible mHA/DRP. **C) Putative mHA:** Each DRP had its nsSNP-encoded peptides filtered by predicted binding to six HLA alleles specific to that DRP. Average number of peptides with binding affinity labeled *presented* (< 500 nM), and *strongly presented* (< 50 nM) is shown.

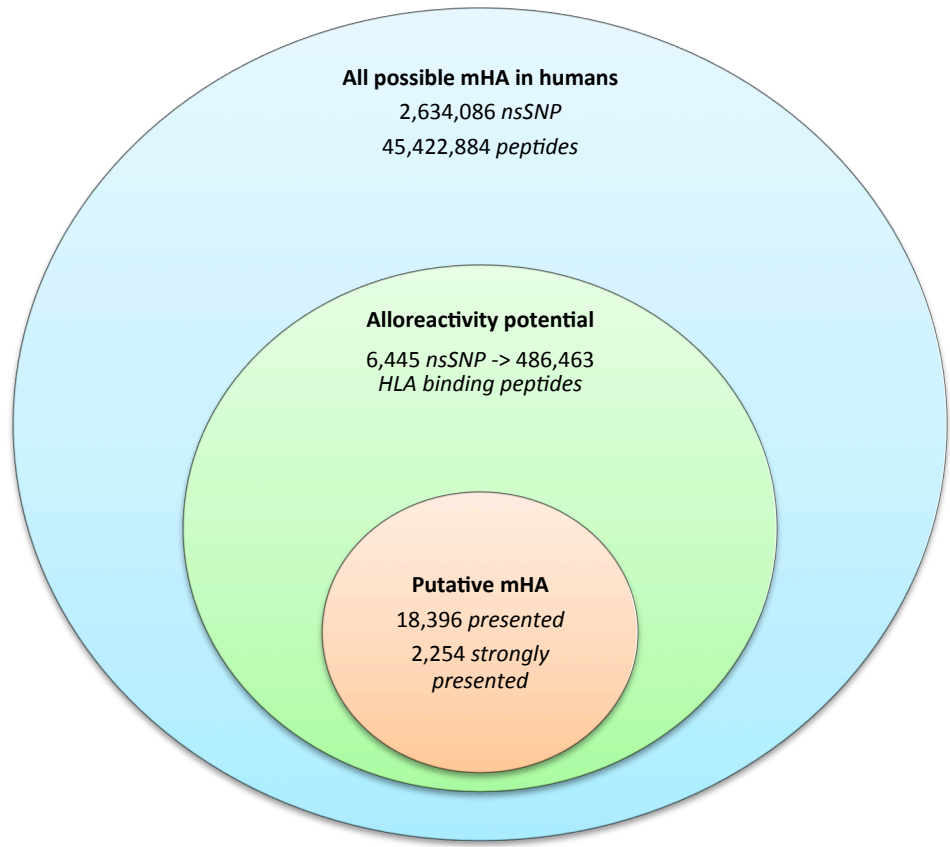

**Figure 3: Whole exome sequence variation and resulting HLA-binding oligopeptides in MRD and URD.**
**A.** Number of nsSNP, and the resulting *presented* (IC50 < 500 nM) and *strongly presented* (IC50 < 50nM) peptides (GVH vector) presented by the HLA in each patient. **B.** Same data as in Figure 3A, presented with the Y-axis changed to Log-scale to better illustrate the SNP to HLA-binding peptide ratio between MRD and URD. Significant difference observed in the distribution of SMM-IEDB predicted presented and strongly presented peptides between MRD and URD. Patients 2, 4, 16, 23 – MRD; 3, 5, 7, 8, 10 – URD SCT recipients.

**A.**

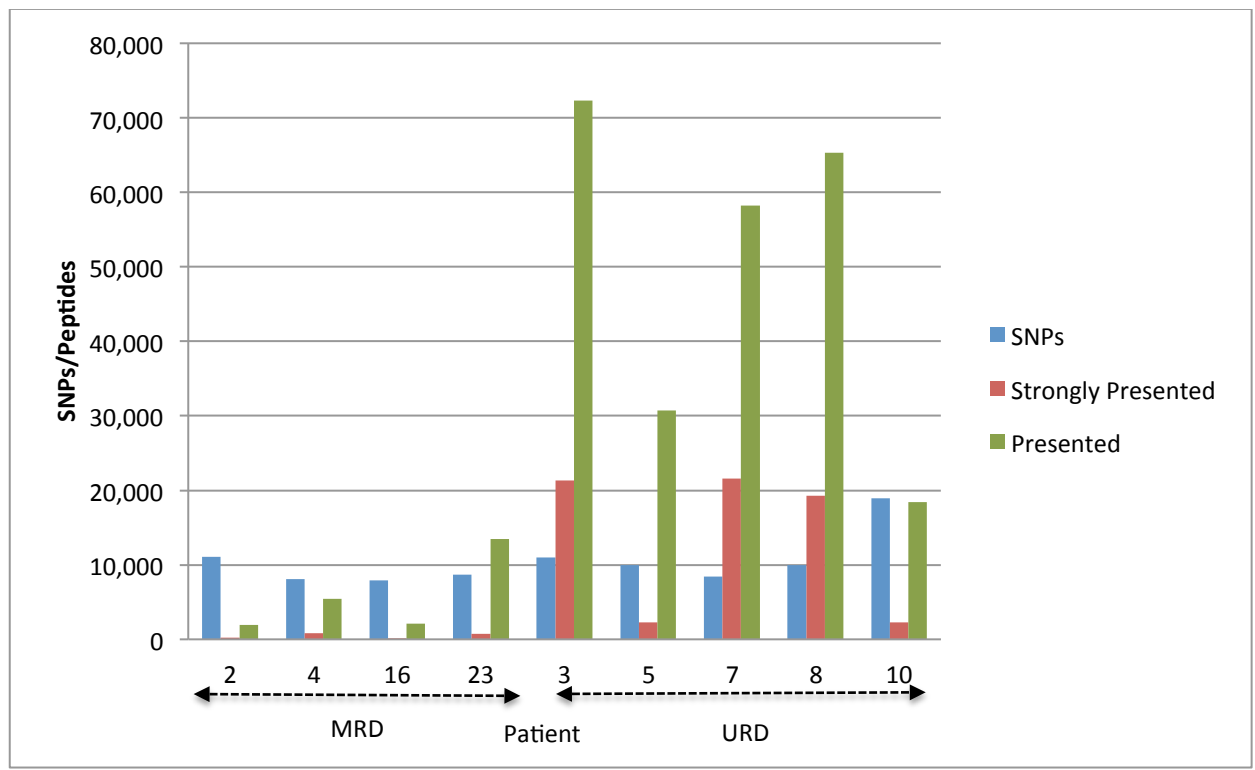

**B.**

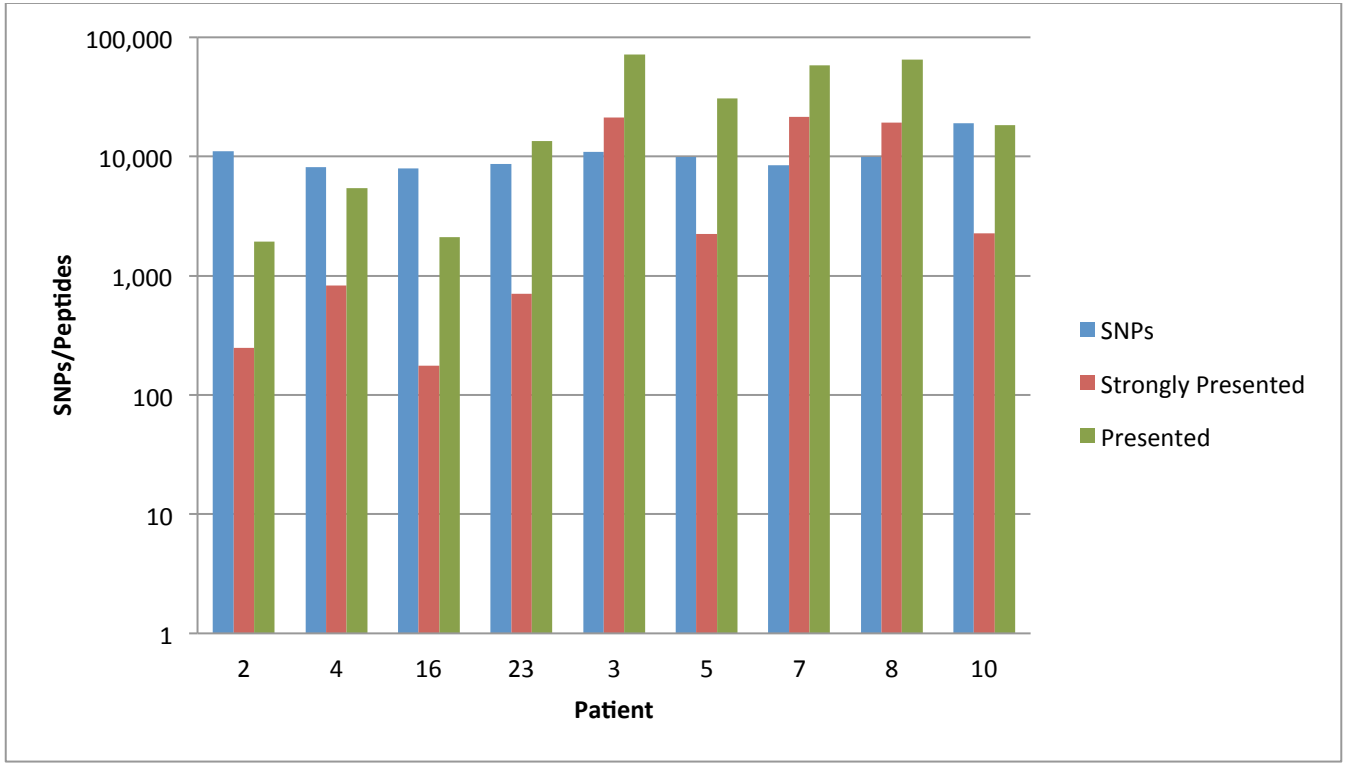

**Figure 4. Peptide-HLA complexes with IC50 values up to 100 nM plotted in descending order of binding affinity.** Depicting difference in the number of and binding affinity (inverse of IC50) of peptide-HLA complexes for each DRP. IC50 distribution is non-linear and described as a polynomial function of the second order, forming a continuum. Marked difference observed between MRD and URD (See Table 2).

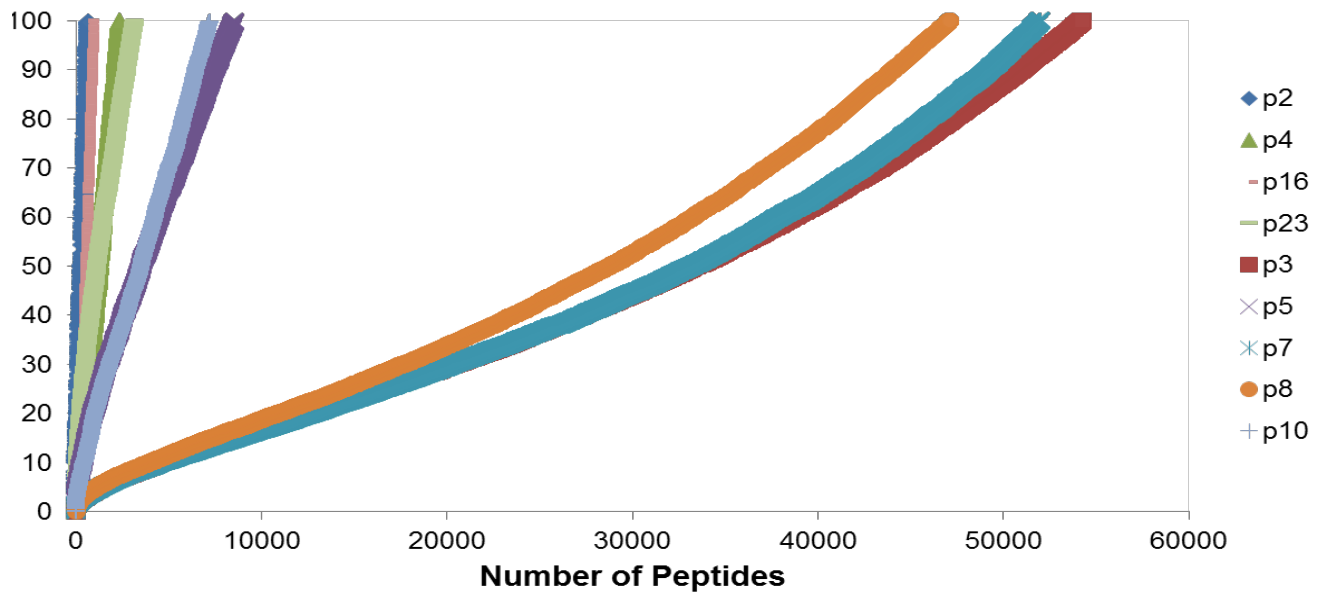

**Figure 5. Unique peptide-HLA complexes (GVH vector) with IC50 <500nM predicted by both SMM and NetMHCpan.** Scatter plots depict the IC50 for unique polymorphic peptide-HLA complexes predicted by the two different algorithms studied. A large number of patient HLA specific strong binding peptides identified by both programs, using SNP data derived from exome sequencing. Only *shared* peptide-HLA complexes predicted to have an IC50 < 500nM by both algorithms included.

Patient 23 (MRD)

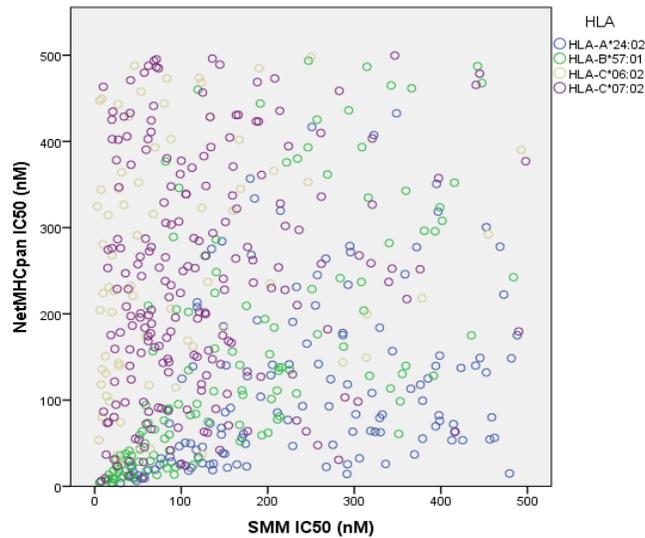

Patient 5 (URD)

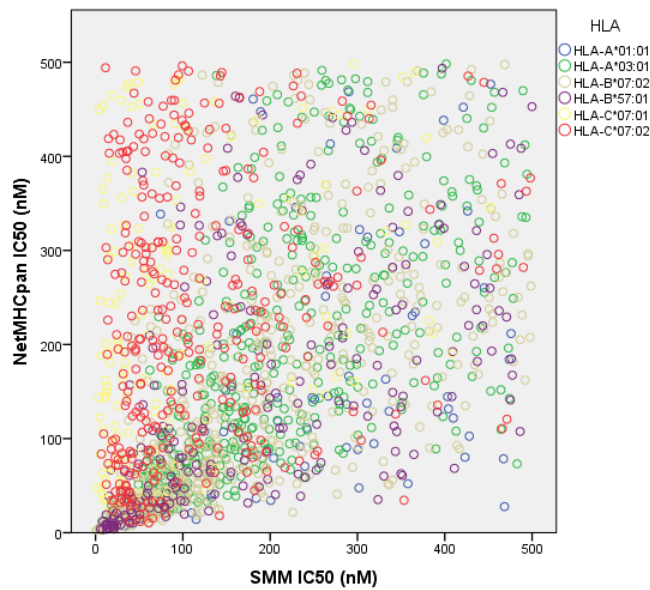

**Figure 6: Tissue Distribution of presented mHA with Gene Expression.** Number of genes coding for mHA (IC50 <500 nM by SMM algorithm) and expressed at a REU >10. European Bioinformatics Institute Illumina Body Map was used to correlate presented peptides with relative gene expression in 16 tissues. Several hundred genes per organ expressed have nsSNP$_{GVH}$, which may generate HLA binding peptides (SMM IEDB data set).

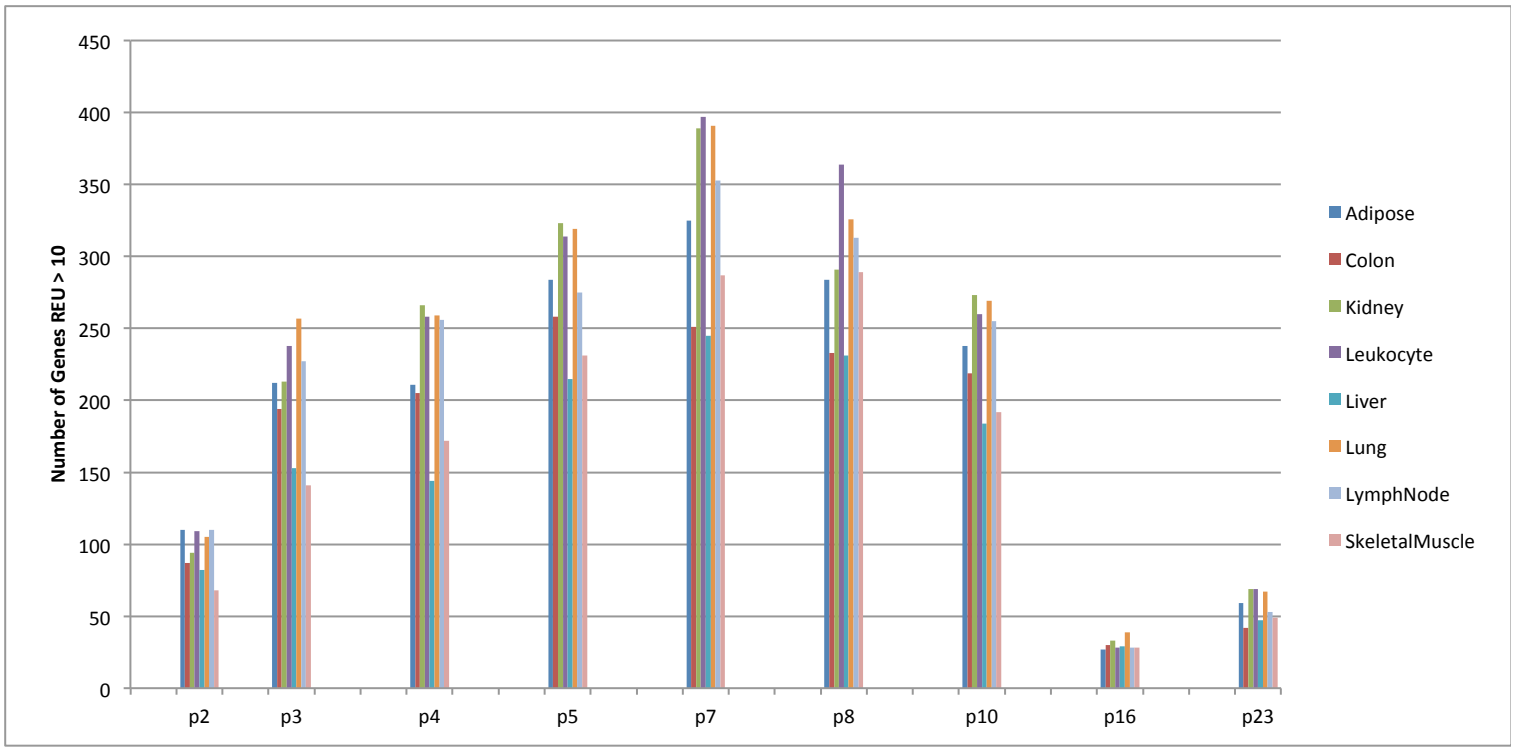

**Figure 7: A quantitative model for the development of GVHD.** Whole exome sequencing identifies all the nsSNP with a GVH vector, yielding a putative *alloreactivity potential* which may be a function (*f*) of the cumulative influence of these polymorphisms. This is represented as a series, listing the sequence of polymorphic exome loci. Substituting individual $_{ns}SNP_{GVH}$ in the equation by peptide-HLA *binding affinity* (reciprocal of IC50) * relative expression level of the gene bearing the $_{ns}SNP_{GVH}$ (for each HLA molecule) yields the HLA-specific alloreactivity potential, in this Re is the relative expression of protein with $_{ns}SNPn_{GVH}$ and resulting peptides (P*n*). In this series, the expression, $\underline{Re_{p1}*(1/IC50_{P1\text{-}HLA\text{-}A1})}$ for each specific peptide-HLA complex, hypothetically represents the T cell clone-specific AP. Multiple peptides constituting this series then drive a proportional oligoclonal T cell expansion in GVHD, as many different mHA are presented by the HLA in an individual, the final distribution conforming to the Power Law. Since T cell clonal expansion in response to presented antigens is influenced by factors such as tissue injury, cytokine milieu and immunosuppression intensity, GVHD likelihood, and its phenotype is determined not only by the ubiquitous mHA, but by the tissue volume and its state (inflammation/injury), and most importantly time at which organ *injury/inflammation* occurs relative to T cell infusion.

Alloreactivity potential $_{GVHD} \approx f\,(_{ns}SNP1_{GVH}+\ _{ns}SNP2_{GVH}+\ldots.\ _{ns}SNPn_{GVH})$

HLA-specific alloreactivity potential $_{GVHD} \approx f\,[(Re_{p1}*(1/IC50_{P1\text{-}HLA\text{-}A1}))+(Re_{p1}*(1/IC50_{P1\text{-}HLA\text{-}A2}))\ldots.$
$+(Re_{pn}*(1/IC50_{Pn\text{-}HLA\text{-}A1}))+(Re_{pn}*(1/IC50_{Pn\text{-}HLA\text{-}A2}))\ldots.]$

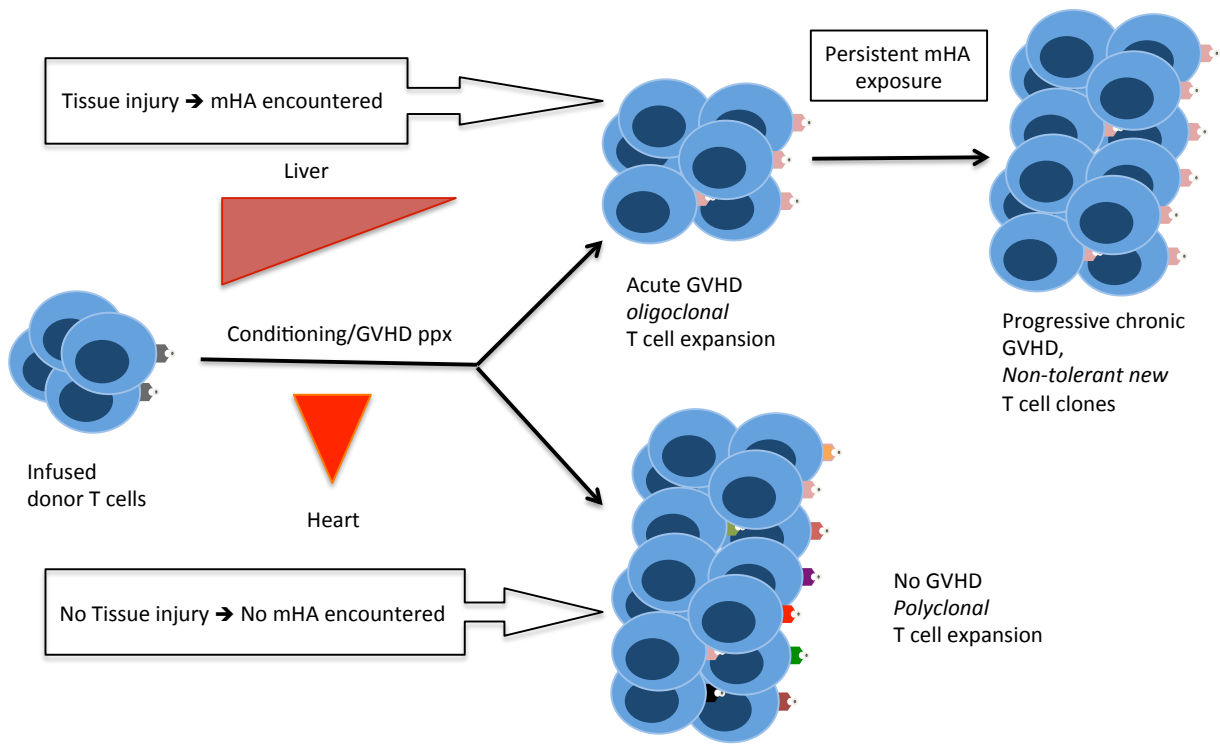

**Supplementary Table 1.** SMM versus MHC binding affinities of various peptides for HLA-B1503 in a single patient (patient 4).

| HLA Allele | Gene Name | Gene ID | Chromosome | Peptide | IC50 SMM | IC50 MHC |
|---|---|---|---|---|---|---|
| HLA-B*15:03 | DSEL | NM_032160 | chr18 | FQWCFYLSF | 0.21 | 2.42 |
| HLA-B*15:03 | KIAA0141 | NM_001142603 | chr5 | FQLSVSITF | 0.69 | 2.02 |
| HLA-B*15:03 | OR5AU1 | NM_001004731 | chr14 | LQRLLFMVF | 0.78 | 3.02 |
| HLA-B*15:03 | DCHS1 | NM_003737 | chr11 | FQRPHYVAF | 0.82 | 1.91 |
| HLA-B*15:03 | SLC22A6 | NM_153278 | chr11 | LQLLVSVPF | 0.86 | 3.40 |
| HLA-B*15:03 | ZNF106 | NM_022473 | chr15 | LQITTCPTF | 1.09 | 2.60 |
| HLA-B*15:03 | OR10P1 | NM_206899 | chr12 | SMMTATIVF | 1.14 | 2.26 |
| HLA-B*15:03 | DNAH14 | NM_001373 | chr1 | SQSKLTSTF | 1.41 | 1.88 |
| HLA-B*15:03 | TARBP1 | NM_005646 | chr1 | MKFGTNAYM | 1.47 | 12.26 |
| HLA-B*15:03 | PREPL | NM_001171603 | chr2 | LKKYHLTKF | 1.48 | 25.40 |
| HLA-B*15:03 | CYP1B1 | NM_000104 | chr2 | LKWPNPENF | 1.53 | 21.21 |
| HLA-B*15:03 | OR2T8 | NM_001005522 | chr1 | LQAVVTLSF | 1.71 | 2.10 |
| HLA-B*15:03 | ASB18 | NM_212556 | chr2 | WQVKSPTTF | 1.72 | 1.97 |
| HLA-B*15:03 | EPX | NM_000502 | chr17 | SQVPLSSAF | 1.75 | 1.85 |
| HLA-B*15:03 | ADAM2 | NM_001464 | chr8 | LMNAIFVSF | 1.75 | 3.10 |
| HLA-B*15:03 | PKP1 | NM_001005337 | chr1 | RHFSSYSQM | 1.98 | 16.71 |
| HLA-B*15:03 | CAPN14 | NM_001145122 | chr2 | RQNEFFTKF | 2.12 | 1.95 |
| HLA-B*15:03 | SH3D19 | NM_001128923 | chr4 | YMHGDVLVM | 2.20 | 2.90 |
| HLA-B*15:03 | ATXN1 | NM_001128164 | chr6 | KMGLSAAPF | 2.42 | 6.55 |
| HLA-B*15:03 | ABP1 | NM_001272072 | chr7 | FAFRLRSSF | 2.49 | 2.77 |
| HLA-B*15:03 | MUC16 | NM_024690 | chr19 | SKHASHSTI | 2.58 | 98.91 |
| HLA-B*15:03 | KIAA0226 | NM_001145642 | chr3 | AKSSSSNLF | 2.68 | 6.05 |
| HLA-B*15:03 | MYO15A | NM_016239 | chr17 | LQVLRAYSF | 2.72 | 3.39 |
| HLA-B*15:03 | SVIL | NM_003174 | chr10 | AKHLWNGSF | 2.74 | 9.85 |
| HLA-B*15:03 | OR8D4 | NM_001005197 | chr11 | RQRHTPMYY | 2.86 | 6.02 |
| HLA-B*15:03 | GPAM | NM_001244949 | chr10 | IMSTHIVAF | 2.93 | 2.55 |
| HLA-B*15:03 | SERPINB4 | NM_002974 | chr18 | RKSKESNIF | 2.97 | 10.77 |
| HLA-B*15:03 | MUC16 | NM_024690 | chr19 | LQSLLGPMF | 3.18 | 4.69 |
| HLA-B*15:03 | OR1S1 | NM_001004458 | chr11 | LKLSCSDTM | 3.27 | 25.72 |
| HLA-B*15:03 | APOBEC3B | NM_001270411 | chr22 | YKCFQLTWF | 3.48 | 37.91 |
| HLA-B*15:03 | C8A | NM_000562 | chr1 | RKAQCGQDF | 3.48 | 16.82 |
| HLA-B*15:03 | OR4D2 | NM_001004707 | chr17 | LQRFLFIMF | 3.52 | 7.07 |

**Supplementary Table 2:** Tissue Gene Expression for Presented Peptides (IC50 <500nM). Number of genes REU for which is >10.

|  | Adipose | Colon | Kidney | Leukocyte | Liver | Lung | Lymph Node | Skeletal Muscle |
|---|---|---|---|---|---|---|---|---|
| P2 | 110 | 87 | 94 | 109 | 82 | 105 | 110 | 68 |
| P3 | 212 | 194 | 213 | 238 | 153 | 257 | 227 | 141 |
| P4 | 211 | 205 | 266 | 258 | 144 | 259 | 256 | 172 |
| P5 | 284 | 258 | 323 | 314 | 215 | 319 | 275 | 231 |
| P7 | 325 | 251 | 389 | 397 | 245 | 391 | 353 | 287 |
| P8 | 284 | 233 | 291 | 364 | 231 | 326 | 313 | 289 |
| P10 | 238 | 219 | 273 | 260 | 184 | 269 | 255 | 192 |
| P16 | 27 | 30 | 33 | 28 | 29 | 39 | 28 | 28 |
| P23 | 59 | 42 | 69 | 69 | 47 | 67 | 53 | 49 |

**Supplementary Figure 1:** Reciprocal of the IC50 up to 100 nM (~binding affinity) (patient 5-URD), when plotted in order of increasing IC50 (analogous to Figure 4), follows the Power Law distribution.

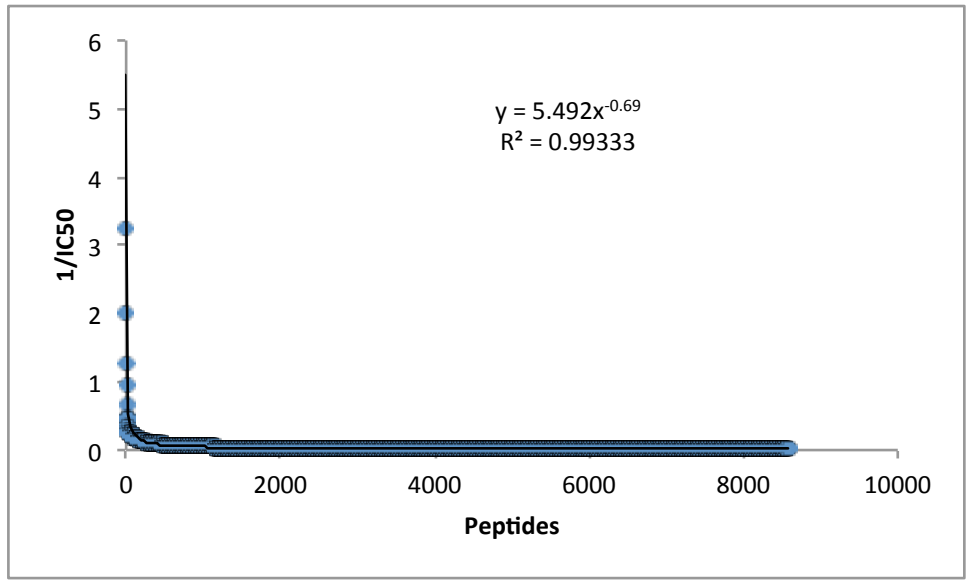

**Supplementary Figure 2**: Number of peptides predicted by the SMM (A) and the NetMHCpan (B) algorithm to be presented by HLA-A, B, and C respectively (IC50 < 500 nM; GVH vector) in MRD (patients 2, 4, 16, 23) and URD (patients 3, 5, 7, 8, 10).

A)

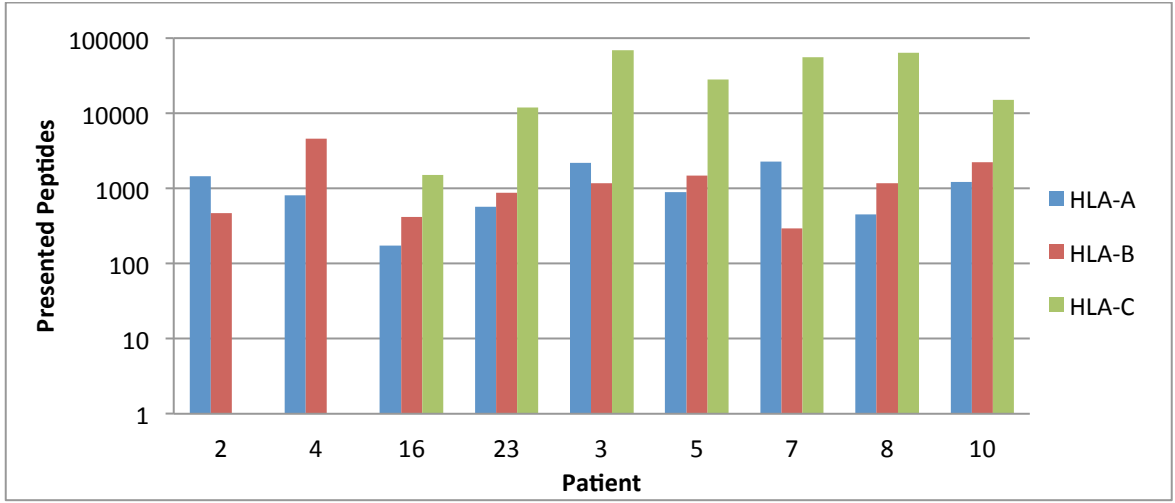

B)

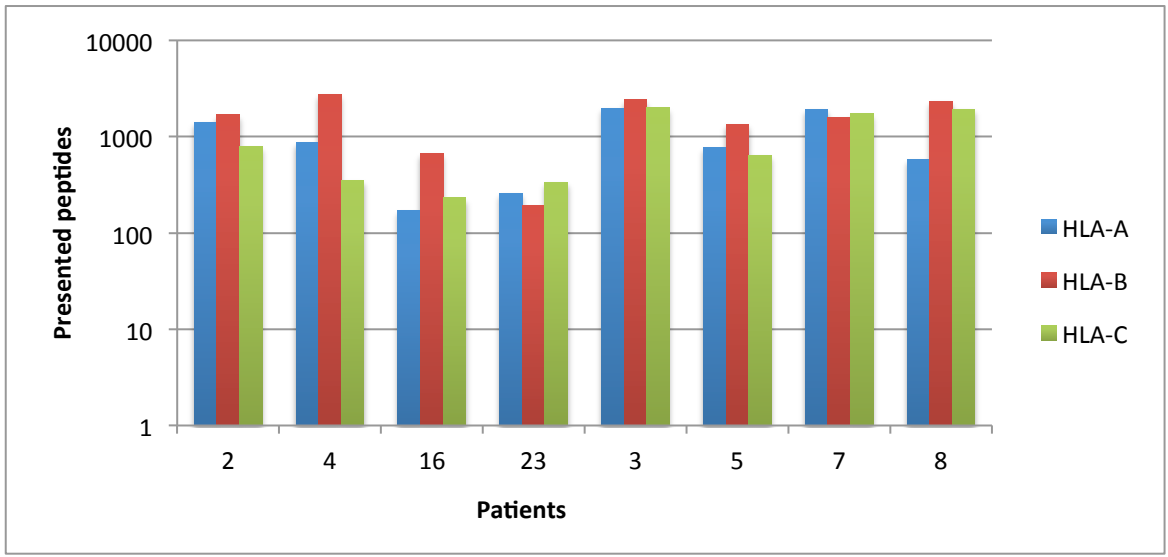

HLA Specific Alloreactivity Potential                                                                                                    37